\documentclass[twocolumn,showpacs,preprintnumbers,amsmath,amssymb]{revtex4}

\usepackage{graphicx}
\usepackage{dcolumn}
\usepackage{bm}

\begin{document}

\preprint{APS/123-QED}

\title{Structural and electronic properties of grain boundaries in graphite:\\Planes of periodically distributed point defects}

\author{J. \v{C}ervenka}
\author{C. F. J. Flipse}%
 \email{c.f.j.flipse@tue.nl}
\affiliation{%
Physics Department, Eindhoven University of Technology, 5600 MB Eindhoven, The Netherlands
}%

\date{\today}

\begin{abstract}
We report on scanning tunneling microscopy and spectroscopy of grain boundaries in highly oriented pyrolytic graphite. Grain boundaries showed a periodic structure and an enhanced charge density compared to the bare graphite surface. Two possible periodic structures have been observed along grain boundaries. A geometrical model producing periodically distributed point defects on the basal plane of graphite has been proposed to explain the structure of grain boundaries. Scanning tunneling spectroscopy on grain boundaries revealed two strong localized states at -0.3~V and 0.4~V.
\end{abstract}

\maketitle

\section{\label{sec:level1}Introduction}

Understanding the defect structures and their role on the electronic structure of graphite is a keystone for carbon nanostructures and carbon materials in general. Defects are inevitable constituents of graphite which have profound influence on its electrical, chemical and other physical properties. Recently, graphene (single layer of graphite) and few-layer graphene showed a number of unconventional properties~\cite{Novoselov1,Novoselov2,Berger} and it seems to be of great importance to understand the influence of defects in this material for possible future applications.

Although graphite is one of the most extensively studied materials there are still new phenomena observed on the graphite surface with scanning tunneling microscope (STM), which are not well understood~\cite{Pong,Niimi1}. In particular, defect structures in the \textit{sp$^2$} bonded carbon lattice have many representations~\cite{Hahn,Hashimoto} and have not been well characterized experimentally yet.

Grain boundaries are one of the most commonly occurring extended defects in highly oriented pyrolytic graphite (HOPG) because of its polycrystalline character. Observations of grain boundaries have been reported on the graphite surface with STM before~\cite{Albrecht,Daulan,Simonis,Pong2} and recently also on few graphene layers grown on C-face of SiC \cite{Varchon}. Periodic structures~\cite{Daulan,Simonis,Pong2,Varchon} and disordered regions~\cite{Albrecht} have been observed along grain boundaries. For a~large angle tilt grain boundary evidence of possible presence of pentagon-heptagon pairs was shown~\cite{Simonis}. Although the structure of various grain boundaries in graphite has been examined with STM, there have not been established a~proper model, which can explain all the observed grain boundaries. Moreover, the electronic structure of grain boundaries has not been investigated so far.

Point defects and extended defects in graphene and graphite have been studied theoretically previously~\cite{Mizes,Pereira,Lehtinen}. In general defects in the carbon honeycomb lattice break the electron-hole symmetry, which leads to formation of localized states at the Fermi energy~\cite{Vozmediano,Peres1}. In the absence of electron-hole symmetry, these states induce a transfer of charge between the defects and the bulk~\cite{Peres1}. Additionally, it has been shown that point defects such as vacancies and hydrogen-terminated vacancies could be magnetic~\cite{Lehtinen,Vozmediano,Yazyev,Wehling,Pisani,Faccio}, showing that electron-electron interactions play an important role in graphene systems because of low electron densities at the Fermi energy. These defects could be of the essential origin of ferromagnetism observed in different graphite samples~\cite{Esquinazi1,Esquinazi2}.

In this paper, we report on an experimental study of grain boundaries in HOPG. Systematic investigation of grain boundaries in graphite has been performed by atomic force microscopy (AFM), scanning tunneling microscopy (STM) and spectroscopy (STS). Grain boundaries showed one-dimensional (1D) superlattices with localized states and enhanced charge density compared to the bare graphite surface. A~crystallographic model producing periodically distributed point defects is introduced to reproduce the observed grain boundaries.

\section{Experimental}
Samples of HOPG of ZYH quality were purchased from NT-MDT. The ZYH quality of HOPG with the mosaic spread 3.5$^\circ$ - 5$^\circ$ has been chosen because it provides a~high population of grain boundaries on the graphite surface. HOPG samples were cleaved by an adhesive tape in air and transferred into a scanning tunneling microscope (Omicron RT and LT STM) working under ultra high vacuum (UHV) condition. The HOPG samples have been heated to 500$^\circ$C in UHV before the STM experiments. STM measurements were performed in the constant current mode with either mechanically formed Pt/Ir tips or electrochemically etched W tips. STS spectra have been obtained by using lock-in amplifier technique. The same samples have been subsequently studied by atomic force microscopy (AFM) using Multimode scanning probe microscope with Nanoscope IV controller from Veeco Instruments in air.

\section{Results and discussions}

\subsection{Structural properties of grain boundaries}

\begin{figure}[t] 
   \centering
  \includegraphics[width=3.3in]{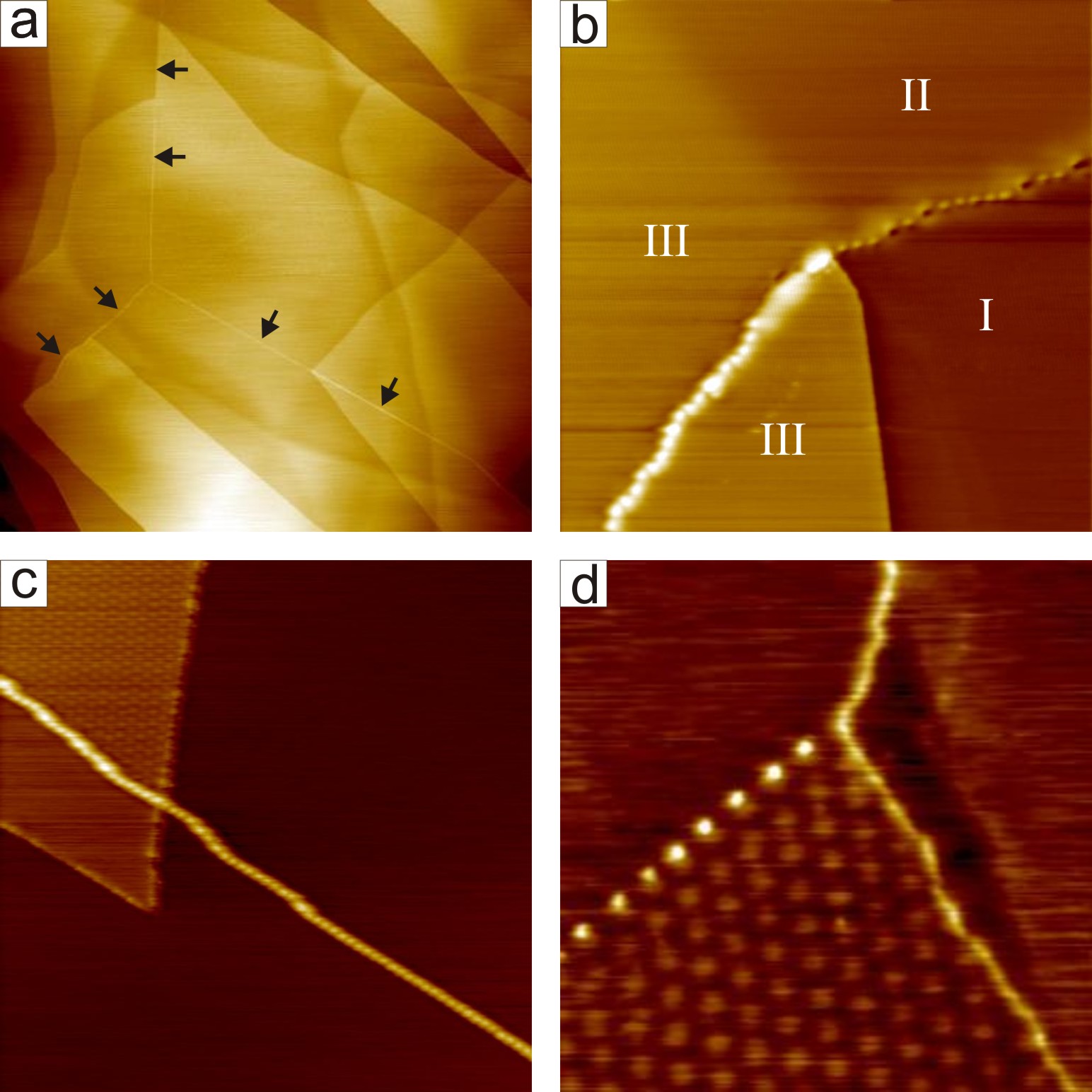}
   \caption{(Color online) (a) AFM image of the HOPG surface with a grain boundary indicated by arrows ($3.5~\times~3.5~\mu\textrm{m}^2$). (b) STM image of a grain boundary continuing as a step edge ($105\times105$~nm$^2$, $U=-0.5$~V, $I=0.5$~nA). (c) STM image of a grain boundary extending over a step edge ($186\times186$~nm$^2$, $U=-0.3$~V, $I=0.3$~nA). (d) STM on grain boundaries bordering a 2D superlattice ($60\times60$~nm$^2$, $U=-0.4$~V, $I~=~0.4$~nA).}
   \label{fig1}
\end{figure}

Figure~\ref{fig1} shows typical examples of grain boundaries observed on the HOPG surface with AFM and STM. In AFM, grain boundaries appear as lines protruding above a graphite surface by a small height up to 0.3~nm. On the other hand in STM, grain boundaries show a periodic one-dimensional superlattice with height corrugations from 0.4~nm up to 1.5~nm, which are almost independent on the applied bias voltage. Since grain boundaries have much smaller height in AFM and the measured corrugation in STM is given by convolution of the topography and the local density of states (DOS) of the substrate, grain boundaries possess enhanced charge density compared to the bare graphite surface. Similar effects of charge accumulation have been observed on defects artificially created by low-energy ions on the graphite surface~\cite{Hahn}. STM images of ion bombarded surfaces showed defects as hillocks, which did not originate from geometric protrusions of a surface as confirmed by AFM but from an increase in DOS near the Fermi energy level~\cite{Hahn}.

Grain boundaries form an continuous network over graphite surface. They interconnect each other as can be seen in figures~\ref{fig1}(a) and~\ref{fig1}(d). AFM and STM images display only the surface signatures of grain boundaries propagating through bulk HOPG crystals. This is demonstrated in figures~\ref{fig1}(a) and~\ref{fig1}(c), where grain boundaries overrun step edges of an arbitrary height without altering their direction, periodicity and corrugation. During the cleavage of the HOPG substrate, grain boundaries pose as weak points, therefore step edges are created out of them on a new formed graphite surface. Figure~\ref{fig1}(b) displays a grain boundary at the bottom left part of the image, which transforms itself into a step edge in the right part of the image. Region I is separated by a monoatomic step (0.35~nm height) from region II and by a double step (0.7~nm height) from region III.

Grain boundaries set bounds to so called 2D superlattices, which are frequently observed on graphite surfaces in STM~\cite{Pong}. Two examples of 2D superattices enclosed by grain boundaries are shown in figures~\ref{fig1}(c) and~\ref{fig1}(d). The most accepted origin of 2D superlattices discussed in the literature is a rotation of the topmost graphite layer with respect to the other layers, which produces Moir\'{e} pattern~\cite{Pong}. Although Moir\'{e} pattern can not explain all the superlattices reported in literature~\cite{Pong} because it does not take into account the interaction between the graphene layers. Nevertheless, it has been in good agreement with all periodicities of 2D superlattices observed in our STM measurements.

\begin{figure}[t] 
   \centering
  \includegraphics[width=3.3in]{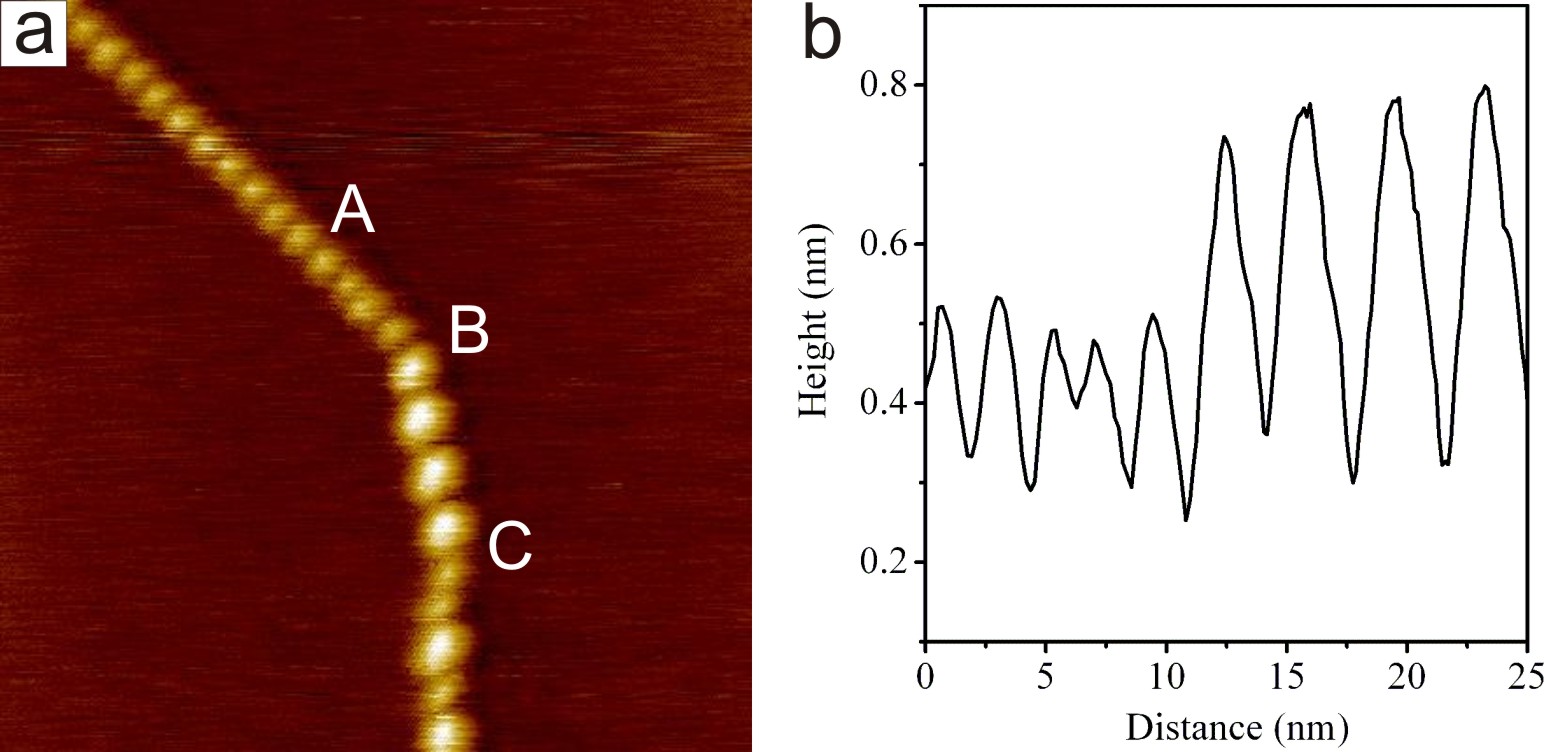}
   \caption{(Color online) (a) STM image of a grain boundary containing two periodicities $D_1=2.18$~nm and $D_2=3.83$~nm. (b) Cross section over the grain boundary in figure (a) along the polyline ABC. Scanning parameters: $50\times50$~nm$^2$, $U~=~1$~V, $I=0.1$~nA.}
   \label{fig2}
\end{figure}

One of the most intriguing properties of grain boundaries is their well defined 1D superlattice periodicity. We have analyzed various grain boundaries on HOPG surfaces. Their superlattice periodicities have been found in the range from 0.5 nm to 10 nm. In principle, two periodicities are observed within a grain boundary as it is demonstrated in figure~\ref{fig2}(a). The second periodicity occurs as the direction of a grain boundary changes by 30$^\circ$ or 90$^\circ$. Figure~\ref{fig2}(b) represents a cross section over the top of the grain boundary from figure~\ref{fig2}(a) going over a polyline ABC with a 30$^\circ$ bend in the point B. The periodicity along the line AB is $D_1=2.18$~nm with a~height corrugation 0.6~nm and the periodicity along the line BC is $D_2=3.83$~nm with a height corrugation of 0.9~nm. The value of the periodicity $D_2$ is approximately $\sqrt{3}D_1$, which will be used as a notation for the second superlattice periodicity later in the text.

\begin{figure}[t] 
   \centering
   \includegraphics[width=3.3in]{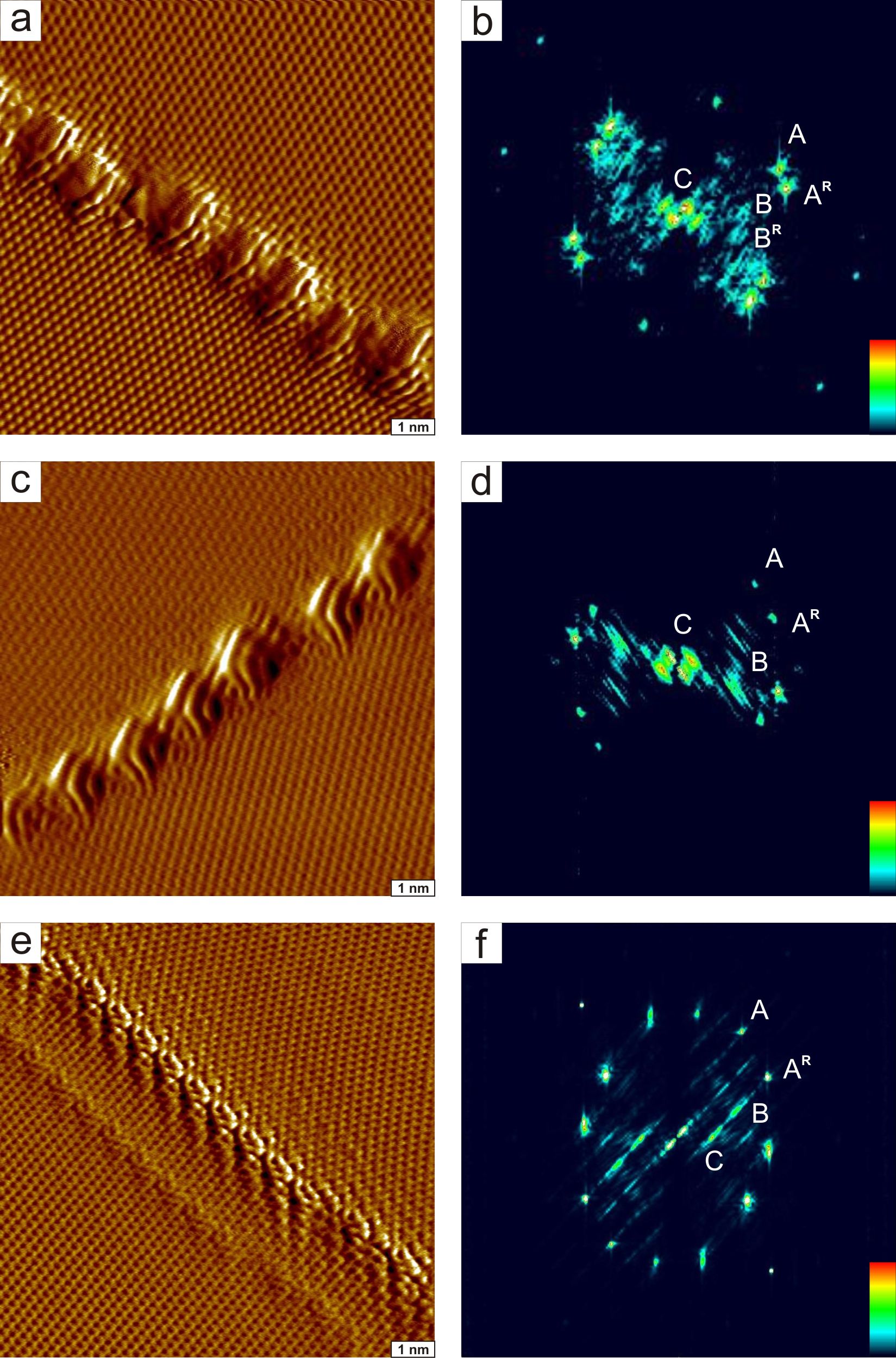}
   \caption{(Color online) Current STM images of three different grain boundaries on HOPG (a), (c) and (e), and their corresponding FFT images (b), (d) and (f), respectively. Grain boundaries show 1D superlattices with periodicities $D=1.25$~nm (a), $\sqrt{3}D=1.4$~nm (c) and $\sqrt{3}D=0.83$~nm (e). The angle between two graphite grains is $\alpha=12^\circ$ (a), $\alpha=18^\circ$ (c) and $\alpha=29.5^\circ$ (e) and the angle between the grain boundary and the graphite lattice is $\beta=25^\circ$ (a), $\beta_{\sqrt{3}D}=9^\circ$ (c) and $\beta_{\sqrt{3}D}=13.5^\circ$ (e). Scanning parameters: $10\times10$ nm$^2$, $U=0.5$~V, $I_t=0.3$~nA.}
   \label{fig3}
\end{figure}

In figure~\ref{fig3}, atomically resolved current STM images of three different grain boundaries and their fast fourier transformation (FFT) images are shown. The grain boundaries exhibit 1D superlattices with periodicities $D=1.25$~nm in figure~\ref{fig3}(a), $\sqrt{3}D=1.4$~nm in figure~\ref{fig3}(c) and $\sqrt{3}D=0.83$~nm in figure~\ref{fig3}(e). It is apparent from these images that grain boundaries in graphite are tilt grain boundaries, which are produced between two rotated graphite grains. No preferential orientation of grains have been found in our measurements. Angles between grains have been determined to be in the interval from $1^\circ$ to $29.5^\circ$. Graphite grains are rotated by angles $12^\circ$, $18^\circ$ and $29.5^\circ$ in figures~\ref{fig3}(a), \ref{fig3}(c) and~\ref{fig3}(e), respectively. The rotation of the graphite grains can be seen as well in the FFT images in figures~\ref{fig3}(b),~\ref{fig3}(d) and~\ref{fig3}(f), where points labeled as A and A$^R$ are forming apexes of two rotated hexagons representing the graphite lattices in the reciprocal space. Six points marked as B demonstrate $\sqrt{3}\times\sqrt{3}R30^{\circ}$ superstructure, which has been observed around point defects and step edges of graphite previously~\cite{Mizes,Kelly}. The $\sqrt{3}\times\sqrt{3}R30^{\circ}$ superstructure is produced by scattering of the free electrons off defects, which generates standing wave patterns in the electron density similar to Friedel oscillations in metals~\cite{Mizes}. The center part of the FFT image marked as C represents the large real space periodicities of the 1D superlattices.

The structure of the grain boundaries can be explained by a simple model, where the superlattice periodicity is determined only by two parameters: $\alpha$ the angle between the grains and $\beta$ the orientation of a grain boundary in respect to the graphite lattice. The orientation toward the graphite lattice can be either $\beta_D=30^\circ~\pm~\alpha/2$ or $\beta_{\sqrt{3}D}=\pm\alpha/2$. The sign depends on the chosen reference direction of the graphite lattice. Two superlattice periodicities could be constructed: $D_1=D$ for $\beta_D$ orientation and $D_2=\sqrt{3}D$ for $\beta_{\sqrt{3}D}$ orientation. The supperlattice periodicity $D$ is given by a simple formula for a~Moir\'{e} pattern $D=d/2sin(\alpha/2)$, where $d=0.246$~nm is the graphite lattice parameter.

\begin{figure}[t] 
   \centering
   \includegraphics[width=3.3in]{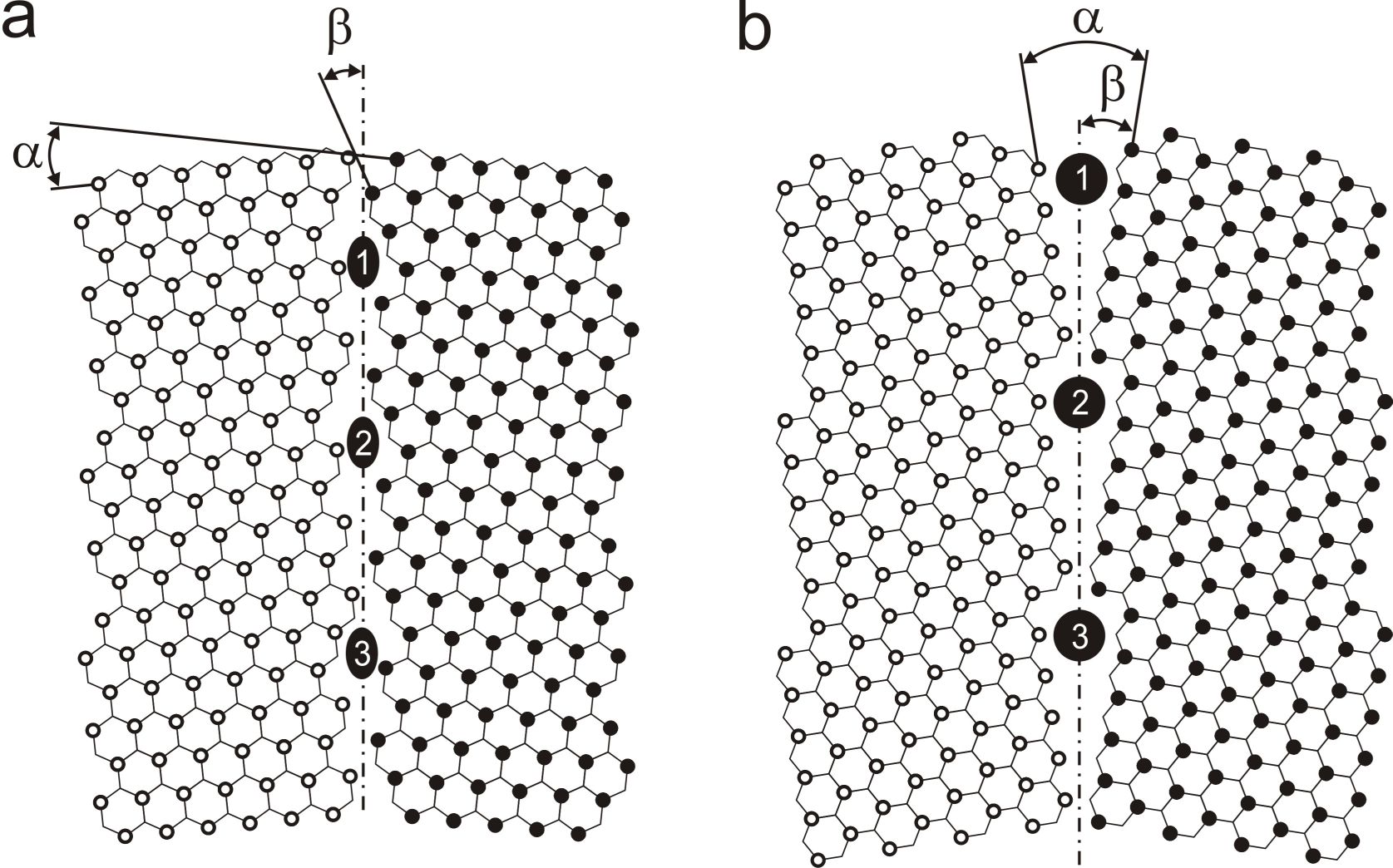}
   \caption{Schematic pictures of grain boundaries in graphite showing two possible superlattice periodicities $D$ (a) and $\sqrt{3}D$ (b). Periodicities within the grain boundaries and angles between the graphite grains have been chosen according to STM observations in figure~\ref{fig3}: $\alpha=12^\circ$, $D=1.18$~nm (a) and $\alpha=18^\circ$, $\sqrt{3}D~=~1.36$~nm (b).}
   \label{fig4}
\end{figure}

In figure~\ref{fig4}, schematic illustrations of the crystallographical structures of two possible orientations of grain boundaries are shown. Periodically distributed point defects are created in this way. They are separated by supperlattice periodicities $D$ in figure~\ref{fig4}(a) and $\sqrt{3}D$ in figure~\ref{fig4}(b). The periodicities of the grain boundaries and angles between the graphite grains have been chosen according to STM observations in figures~\ref{fig3}(a) and \ref{fig3}(b). The combination of these two models within a grain boundary describes well all the possible internal structures and orientations of grain boundaries observed on the graphite surfaces. In addition, grain boundaries with a large angle tilt such as shown in figure~\ref{fig3}(e) would produce row of defects, which are resembling a structure with repeating pentagon-heptagon pairs similarly as was proposed by Simonis et al.~\cite{Simonis}.

Grain boundaries have two basic shapes similarly like graphite edges, which are rotated by $30^\circ$ towards each other. The orientation $\beta_D$ in figure~\ref{fig3}(a) has an armchair character at the axis of the grain boundary, while the $\beta_{\sqrt{3}D}$ orientation in figure~\ref{fig3}(b) has a zigzag character. As it was mentioned before, grain boundaries are weak spots of graphite lattice, therefore step edges are produced out of them during the cleavage. If an edge would be created from the grain boundary by cutting it into half, the edges would have segments of zigzag or armchair edge of the maximum length as the superlattice periodicity $D$ or $\sqrt{3}D$. Previous STM studies of step edges on graphite have found a short length of zigzag edges (up to 2~nm) alternated by armchair segments, while the energetically more stable armchair edges had lengths up to hundred nanometers~\cite{Kobayashi2}. The observed periodicities of the grain boundaries have been found in the same range between 0.5 to 10~nm. This could indicate that short alternating zigzag and armchair edges are created out of grain boundaries.

\subsection{Electronic properties of grain boundaries}

\begin{figure}[b] 
   \centering
   \includegraphics[width=2.8in]{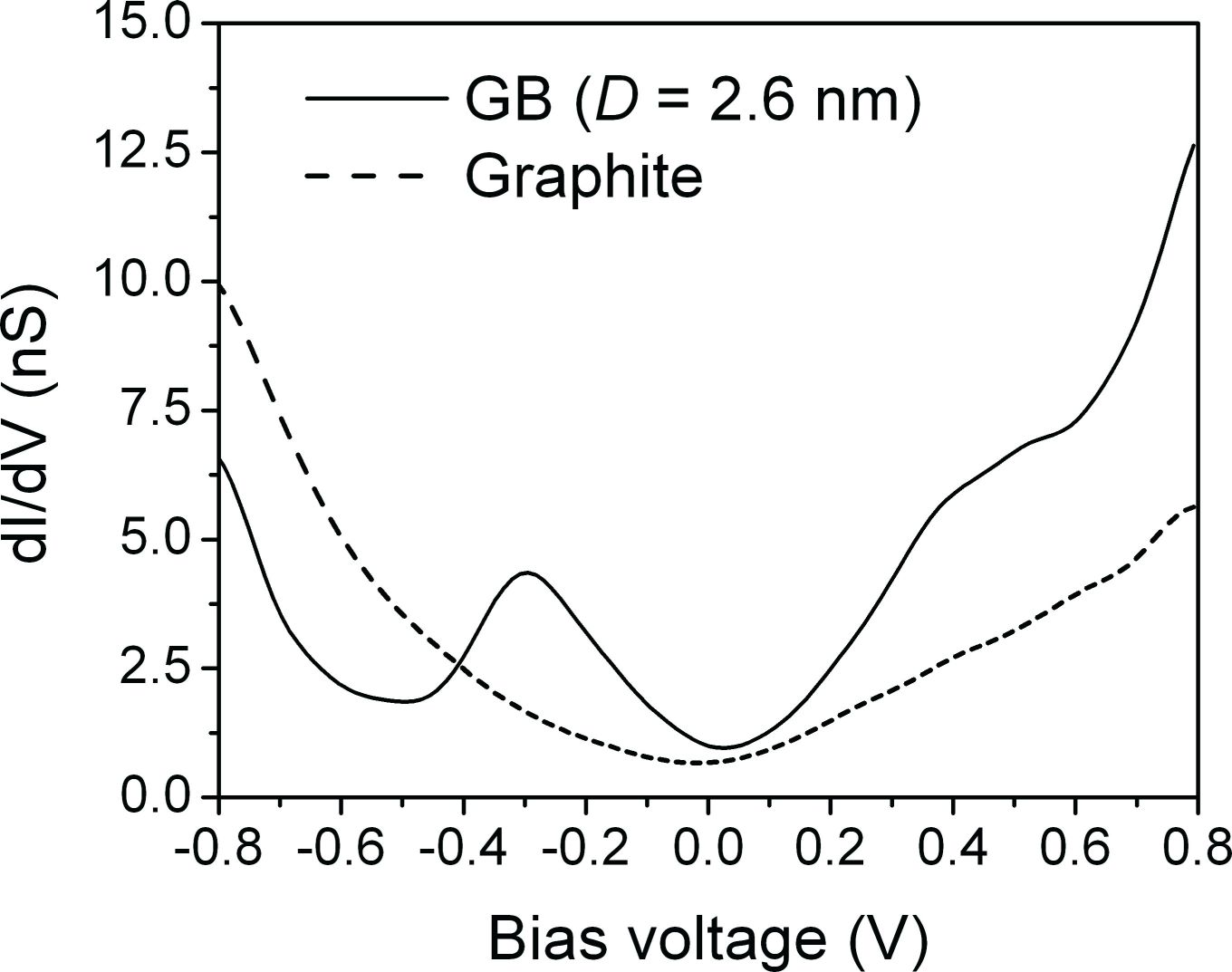}
   \caption{$dI/dV$ curves measured on a grain boundary (GB) with $D~=~2.6~$nm and on the bare graphite surface at room temperature ($U=-0.5$~V, $I~=~0.5$~nA). Two localized states at -0.3~V and 0.4~V are observed on the grain boundary.}
   \label{fig5}
\end{figure}

Scanning tunneling spectroscopy has been measured on grain boundaries and on a clean graphite surface for comparison. In figure~\ref{fig5}, two $dI/dV$ spectra measured on the top of a grain boundary and on the clean graphite surface are shown. STS curve measured on the top of the grain boundaries with the superlattice periodicity $D~=~2.6~$nm exhibits two strong localized states, which are not seen on the clean graphite surface. The positions of the localized states are -0.3~V and 0.4~V.

Various point defects in graphene and graphite have been studied theoretically before~\cite{Mizes,Pereira,Lehtinen,Vozmediano,Peres1,Yazyev,Wehling}. In general, by introducing a defect in the carbon honeycomb lattice the electron-hole symmetry is broken, which leads to creation of localized state at the Fermi energy and to the phenomenon of self-doping~\cite{Vozmediano,Peres1}. The self-doping denotes charge transfer to/from defects to the bulk, which is in accordance with an increased charge DOS at the grain boundaries as was mentioned before.

Since graphene systems have low electron densities at the Fermi energy, electron-electron interactions play an important role as recent experiments showed~\cite{Gruneis}. In the presence of a local repulsive electron-electron interaction the localized states will become polarized, leading to the formation of local moments~\cite{Vozmediano}. This has been illustrated in DFT studies of point defects in graphite such as vacancies and hydrogen-terminated vacancies. These defects revealed to be magnetic having a local magnetic moment larger than 1$\mu_B$~\cite{Lehtinen,Yazyev}. Spin polarized DOS of these systems showed two strongly spin polarized impurity states in the empty and filled states~\cite{Yazyev,Wehling}. The role of different distances between single vacancies has been studied in the DFT study of an 3D array of single vacancies in graphite~\cite{Faccio}. Different sizes of supercells containing single vacancies have been constructed for this purpose. Two spin polarized states have been formed for small supercells, supporting ferrimagnetic order up to the distance 1 nm among the vacancies~\cite{Faccio}. The $5\times5\times1$ supercell (1.23 nm separated vacancies) did not show a net magnetic moment in graphite and a single localized peak around Fermi energy has been observed in spin polarized LDOS. In graphene, the $5\times5$ supercell exhibited still a~net magnetic moment of 1.72$\mu_B$~\cite{Faccio}. Our results show very similar results compared to the theoretical predictions of Faccio \textit{et al.}~\cite{Faccio}, where two split localized states are created. Nevertheless, it cannot be concluded only from STM and STS measurements that grain boundaries are magnetic and other technique like spin polarized STM has to be done in order to prove it.

Another origin of two split localized states in the grain boundaries could be the internal structure of point defects within grain boundaries. Point defects in graphite can exist in several forms, such as single and multiple vacancies, intersticials, Stone Wales defects and other more complicated point defects. All of them can essentially occur in grain boundaries. Moreover, they can be saturated by different atoms like hydrogen, oxygen or nitrogen. Using only STM, it does not allow us to extract the exact atomic structure of defects. Nevertheless, the structure of defects is reflected in the shape and the symmetry of the charge modulation around the defects as it has been shown in a theoretical study of a single atomic and double atomic defects in graphene~\cite{Wehling}. The single atomic defect resulted in a simple trigonal symmetry in the charge modulation around the defect, while double atomic defect demonstrated two fold symmetry. From this point of view, grain boundaries contain more complicated point defects as seen in figure~\ref{fig3}. In order to discern between the two proposed possibilities for diverse DOS of grain boundaries an appropriate calculation has to be done, which is going to be difficult especially for grain boundaries with large periodicities.

\section{Conclusions}

In conclusion, a systematic scanning tunneling microscopy and spectroscopy study of grain boundaries in highly oriented pyrolytic graphite have been done. Different grain boundary geometries have been characterized with a focus on their electronic structure. Grain boundaries showed a periodic structure and an enhanced charge density compared to the bare graphite surface. Two possible periodic structures has been observed along grain boundaries. A geometrical model producing periodically distributed point defects on the basal plane of graphite has been proposed to explain the structure of grain boundaries. Scanning tunneling spectroscopy on grain boundaries revealed two strong localized states at -0.3~V and 0.4~V.

\begin{acknowledgments}
This research was supported by Nanoned.
\end{acknowledgments}


\begin{thebibliography}{100}

\bibitem{Novoselov1} K. S. Novoselov, A. K. Geim, S. V. Morozov, D. Jiang, Y. Zhang, S. V. Dubonos, I. V. Grigorieva, A. A. Firsov, \textit{Science} \textbf{306}, 666 (2004).
\bibitem{Novoselov2} K. S. Novoselov, A. K. Geim, S. V. Morozov, D. Jiang, M.I. Katsnelson, I. V. Grigorieva S. V. Dubonos, A. A. Firsov, \textit{Nature} \textbf{438}, 197 (2005).
\bibitem{Berger} C. Berger, Z. Song, T. Li, X. Li, X. Wu, N. Brown, C. Naud, D. Mayou, A. N. Marchenkov, E.H. Conrad, P. N. First, and W. A. de Heer, \textit{Science} \textbf{312}, 1191 (2006).

\bibitem{Pong} W. T. Pong and C. Durkan, \textit{J. Phys. D} \textbf{38}, R329 (2005).
\bibitem{Niimi1} Y. Niimi, H. Kambara, T. Matsui, D. Yoshioka, and H. Fukuyama, \textit{Phys. Rev. Lett.} \textbf{97}, 236804 (2006).

\bibitem{Hahn} J. R. Hahn and H. Kang, \textit{Phys. Rev. B} \textbf{60}, 6007 (1999).
\bibitem{Hashimoto} A. Hashimoto, K. Suenaga, A. Gloter, K. Urita, and S. Iijima, \textit{Nature} \textbf{430}, 870 (2004).

\bibitem{Albrecht} T. R. Albrecht, H. A. Mizes, J. Nogami, Sang-il Park, and C. F. Quate, \textit{Appl. Phys. Lett.} \textbf{52}, 362, (1988).
\bibitem{Daulan} C. Daulan, A. Derr\'{e}, S. Flandrois, J. C. Roux, and H. Saadaoui, \textit{J. Phys. I France} \textbf{5}, 1111 (1995).
\bibitem{Simonis} P. Simonis, C. Goffaux, P. A. Thiry, L. P. Biro, Ph. Lambin, V Meunier, \textit{Surf. Sci.} \textbf{511}, 319 (2002).
\bibitem{Pong2} W. T. Pong, J. Bendall, C. Durkan, \textit{Surf. Sci.} \textbf{601}, 498 (2007).
\bibitem{Varchon} F. Varchon, P. Mallet, L. Magaud, and J. Y. Veuillen, \textit{Phys. Rev. B} \textbf{77}, 165415 (2008).

\bibitem{Mizes} H. A. Mizes and J. S. Foster, \textit{Science} \textbf{244}, 559 (1989).
\bibitem{Kelly} K. F. Kelly, D. Sarkar, G. D. Hale, S. J. Oldenburg and N. J. Halas, {\it Science} {\bf 273}, 1371 (1996).
\bibitem{Pereira} V. M. Pereira, F. Guinea, J. M. B. Lopes dos Santos, N. M. R. Peres, and A. H. Castro Neto, \textit{Phys. Rev. Lett.} \textbf{96}, 036801 (2006).
\bibitem{Peres1}  N. M. R. Peres, F. Guinea, and A. H. C. Neto, \textit{Phys. Rev. B} \textbf{73}, 125411 (2006).

\bibitem{Vozmediano}  M. A. H. Vozmediano, M. P. Lopez-Sancho, T. Stauber, and F. Guinea, \textit{Phys. Rev. B} \textbf{72}, 155121 (2005).
\bibitem{Lehtinen} P. O. Lehtinen et al., \textit{Phys. Rev. Lett.} \textbf{93}, 187202 (2004).
\bibitem{Yazyev} O. V. Yazyev and L. Helm, \textit{Phys. Rev. B} \textbf{75}, 125408 (2007).
\bibitem{Wehling}   T. O. Wehling, A. V. Balatsky, M. I. Katsnelson, A. I. Lichtenstein, K. Scharnberg, and R. Wiesendanger, \textit{Phys. Rev. B} \textbf{75}, 125425 (2007).
\bibitem{Pisani} L. Pisani, B. Montanari, and N. M. Harrison, \textit{New Journal of Physics} \textbf{10}, 033002 (2008).
\bibitem{Faccio} R. Faccio, H. Pardo, P. A. Denis, R. Yoshikawa Oeiras, F. M. Ara\'{u}jo-Moreira, M. Ver\'{\i}ssimo-Alves, and A. W. Mombr\'{u}, \textit{Phys. Rev. B} \textbf{77}, 035416 (2008).


\bibitem{Esquinazi1} P. Esquinazi, A. Setzer, R. H\"{o}hne, C. Semmelhack, Y. Kopelevich, D. Spemann, T. Butz, B. Kohlstrunk and M. Lösche, \textit{Phys. Rev. B} \textbf{66}, 024429 (2002).
\bibitem{Esquinazi2} P. Esquinazi, D. Spermann, R. H\"{o}hne, A. Setzer, K.-H. Han, and T. Butz, \textit{Phys. Rev. Lett.} \textbf{91}, 227201 (2003).

\bibitem{Kobayashi2}   Y. Kobayashi \textit{et al.}, \textit{Phys. Rev. B} \textbf{73}, 125415 (2006).
\bibitem{Gruneis} A. Gr\"{u}neis \textit{et al.}, \textit{Phys. Rev. Lett.} \textbf{100}, 037601 (2008).

\end{thebibliography}
\end{document}